\documentclass[%
 reprint,
 aip,
 amsmath,amssymb,
 aps,
]{revtex4-2}
\usepackage{graphicx}
\usepackage{placeins}
\usepackage{dcolumn}
\usepackage{bm}
\usepackage{caption}
\usepackage{subcaption}
\usepackage{hyperref}
\usepackage{xcolor} 
\usepackage[normalem]{ulem} 	
\newcommand{\blue}[1]{\textcolor{blue}{#1}}
\newcommand{\gray}[1]{\textcolor{gray}{#1}}

\newcommand{\discuss}[1]{\textcolor{orange}{#1}}

\newcommand{\phib}[0]{\ensuremath{\Phi_\mathrm{B}}}

\newcommand{\fq}[0]{\ensuremath{f_q}}
\newcommand{\fr}[0]{\ensuremath{f_\mathrm{res}}}
\newcommand{\fpr}[0]{\ensuremath{f_\mathrm{probe}}}
\newcommand{\etal}[0]{\textit{et al.}}

\begin{document}

\title{Gradiometric, Fully Tunable C-Shunted Flux Qubits}

\author{B. Berlitz (corresponding author)}\email{benedikt.berlitz@kit.edu}
 \affiliation{Physikalisches Institut, Karlsruhe Institute of Technology (KIT), Karlsruhe, Germany}
 \author{A. Händel}
 \affiliation{Physikalisches Institut, Karlsruhe Institute of Technology (KIT), Karlsruhe, Germany}
\author{E. Daum}
 \affiliation{Physikalisches Institut, Karlsruhe Institute of Technology (KIT), Karlsruhe, Germany}
\author{A.V. Ustinov}
 \affiliation{Physikalisches Institut, Karlsruhe Institute of Technology (KIT), Karlsruhe, Germany}
 \affiliation{Institute for Quantum Materials, Karlsruhe Institute of Technology (KIT), Karlsruhe, Germany}
\author{J. Lisenfeld}
 \affiliation{Physikalisches Institut, Karlsruhe Institute of Technology (KIT), Karlsruhe, Germany}

\date{\today}


\begin{abstract}
Fully tunable flux qubits offer in-situ  and independent controls of their energy potential asymmetry and tunnel barrier, making them versatile tools for quantum computation and the study of decoherence sources. However, only short coherence times have been demonstrated so far with this type of qubit.
Here, we present a capacitively shunted flux qubit featuring improved relaxation times up to  $T_1=25$\,\textmu s and a frequency tunability range of $\sim$20 GHz at the flux-insensitive sweet spot.
As a model application, we demonstrate detection of two-level tunneling defects in a frequency range spanning one octave.
\end{abstract}

\maketitle

\section{\label{sec:intro}Introduction}
\begin{figure}[t]
\centering
\includegraphics[width=\columnwidth]{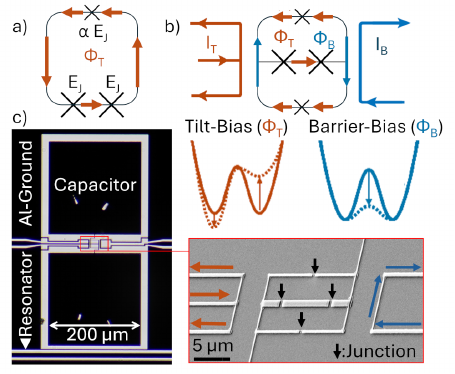}
\caption{a) Idealized circuit diagram of a 3-Junction flux qubit, with one junction having a lower critical current by a factor of $\alpha$. b) Circuit diagram of a gradiometric fully tunable flux qubit. The currents ($I_T,I_B$) in the local bias lines induce the fluxes $\Phi_T$ and $\Phi_B$, providing control of the potential asymmetry and barrier-height, respectively, which is illustrated by the orange and blue qubit potentials. c) Micrograph of a fully tunable C-shunted flux qubit (sample A), with an inset zooming in on the junction layout under a 45° angle (corresponding area of sample B).}
\label{fig:sample}
\end{figure}
\begin{figure*}[t]
\centering
\includegraphics[width=\textwidth]{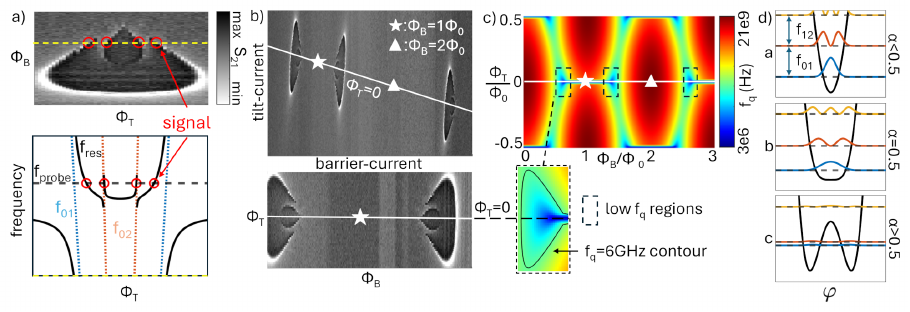}
\caption{a) principle of a $\Phi_T$-$\Phi_B$-sweep calibration measurement. A signal (red circles) is detected, when \fr~ is shifted into resonance with \fpr~ by a qubit transition. b) Calibration measurements used to identify symmetry points in the $\Phi_T$-$\Phi_B$-landscape and the bias line cross-talk (corrected in the bottom measurement). c) Numerical calculation of the qubit resonance frequency $f_\mathrm{q}$, spanning from 3 MHz to 21 GHz. d) Numerical calculation of the qubit potential for different $\alpha$-values, including the 3 lowest qubit states. Shown measurements were performed on sample A.}
\label{fig:results}
\end{figure*}

Superconducting micro-circuits have become a tremendous testbed to explore quantum coherence in electrically controlled solid-state systems.
There are various types of superconducting qubits such as the charge, phase, and flux qubit, which differ by the degree of freedom that dominates their energy. While not as widely used as transmon qubits in current large-scale architectures\cite{WillowProcessor,SycamoreProcessor}, flux qubits, which offer a higher anharmonicity, have served as an important tool for research in superconducting quantum circuits over the past two decades.\\
The original flux qubit\cite{Orlando1999,Mooij1999} consisted of a superconducting loop that is interrupted by three Josephson junctions, one of which has smaller critical current than the other two by a factor $0.5 <\alpha <1$, which defines the Josephson energies $E_J$ for one large junction and $\alpha E_J$ for the small junction (see Fig.\ref{fig:sample}a). It features a double-well potential whose asymmetry is controlled by an external flux $\Phi_T$ (see Fig.\ref{fig:sample}b). In early experiments, flux qubits were read out by measuring the flux through the qubit loop using a DC-SQUID in switching-current\cite{switchingcurrentsChiorescu_2003} or dispersive measurements\cite{dispersiveLupa}, and were used to demonstrate two-qubit gates\cite{de2010selective}, access the ultra-strong coupling cavity QED regime\cite{strongcoupling_Forn_Diaz,strongcouplingNiemczyk2010}, and multiplexed qubit readout\cite{Multiplexed_Jerger2012}.\\
The quantum coherence of flux qubits steadily improved over the past years. 
Bertet \etal\cite{T1_Chiorescu} and Yoshihara \etal\cite{T1_Yoshihara} reported $T_1$ times of 2-4~\textmu s in early experiments, which were likely limited by strongly coupled readout-SQUIDs. Subsequently, longer $T_1$ times of 6-20~\textmu s were reached by Orgiazzi \etal\cite{Orgiazzi2016} and Stern \etal\cite{CavityStern}, by replacing the readout-SQUID with coplanar waveguide resonators and 3D cavities, respectively.\\
Another design improvement was the addition of a shunt-capacitor. While in early experiments using interdigitated capacitors by Steffen \etal\cite{CapacitanceSteffen} and Córcoles \etal\cite{CapacitanceCorcoles}, $T_1$ times were limited to 1-6~\textmu s, Yan \etal~ demonstrated coherence times of $T_1 \approx 50$ \textmu s, using larger square-plate shunt-capacitors\cite{Yan2016}. With this breakthrough, which they attributed to reduced dielectric loss in their shunt capacitors and a  reduction in the qubit persistent current, flux qubits reached similar coherence times as transmon qubits. \\
For the symmetric double-well potential at a magnetic flux bias of $\Phi_T = \Phi_0 / 2$, a flux qubit has minimum resonance frequency \fq~ and lowest dephasing due to flux noise being suppressed in first order. Phase coherence ($T_2$) quickly degrades away from the optimal flux bias point, and this severely limits the practically useful qubit tunability range.
In a "fully tunable" or "gap-tunable" flux qubit, this is addressed by replacing the small junction with a DC-SQUID, thereby making the effective junction area ratio $\alpha$ tunable \textit{in-situ} by an additional control flux $\Phi_B$. This was first shown by Paauw \etal \cite{PaauwGapTunable}. 
Soon after, Poletto \etal~demonstrated operation in both the double-well flux qubit ($0.5 < \alpha < 1$) and the single-well phase qubit ($\alpha<0.5$) regime \cite{poletto2009tunability}. This control of the qubit potential makes gap-tunable flux qubits attractive for different applications, such as quantum annealing\cite{AnnealingJohnson2011,Saida2022}, as couplers and computational qubits in quantum processors\cite{Zhu2010,Chang2023}, for use in quantum metamaterials\cite{Shulga2018} and for microwave-free qubit manipulation \cite{poletto2009coherent}. Gap-tunable flux qubits are particularly well suited for investigating decoherence sources, since a single device can access both flux-like and phase-like operating regimes. Their wide tunability enhances the accessible range for coherence spectroscopy, where qubit relaxation serves as a probe of its electromagnetic environment to reveal interactions with parasitic circuit modes or with microscopic two-level-defects (TLS). \\
So far, balancing the enhanced qubit tunability with good qubit coherence has been proven to be difficult. Early gap-tunable designs exhibited $T_1$ times between 1.4 ns and 1.5 $\mu$s, which have been attributed to thermal noise\cite{Paauw2009}, and dielectric loss\cite{poletto2009coherent}, respectively. In a recent experiment using an asymmetric $\alpha$-junction SQUID, Chang \etal~reported $T_1$ times up to 8 $\mu$s \cite{Chang2023}, attributing this limit to dielectric loss. However, this approach limits the tuning range of the device (about 7 GHz was reported). Despite this significant improvement in energy relaxation, a gap-tunable flux qubit, which combines access to a wide $\alpha$-range with a long $T_1$ time has not yet been achieved.\\
Here, we present a gap-tunable capacitively shunted flux qubit that is based on the design of Yan \etal\cite{Yan2016}, which combines the coherence of the C-shunted flux qubit with wide tunability and \textit{in-situ} $\alpha$-control. Its half-gradiometric design provides independent local control of barrier and asymmetry-flux biases. We observed $T_1$ times up to 25 $\mu s$ at 3.5 GHz, which corresponds to a qubit quality factor of $2\pi\fq\cdot T_1 \approx$ 500k. Furthermore, we demonstrate strain-tuned TLS spectroscopy as a model application. This technique, previously implemented with phase and transmon qubits\cite{Lisenfeld2016,Lisenfeld2019}, demonstrates the suitability of our device for investigating microscopic defects in a wide frequency range. Taking advantage of this feature in future experiments could provide deeper insights into fundamental material properties such as the frequency dependence of the TLS state density, which is a central ingredient in atomic defect models\cite{Mueller2019}.
\\

\section{\label{sec:design}method}
\begin{figure*}[t]
\centering
\includegraphics[width=2\columnwidth]{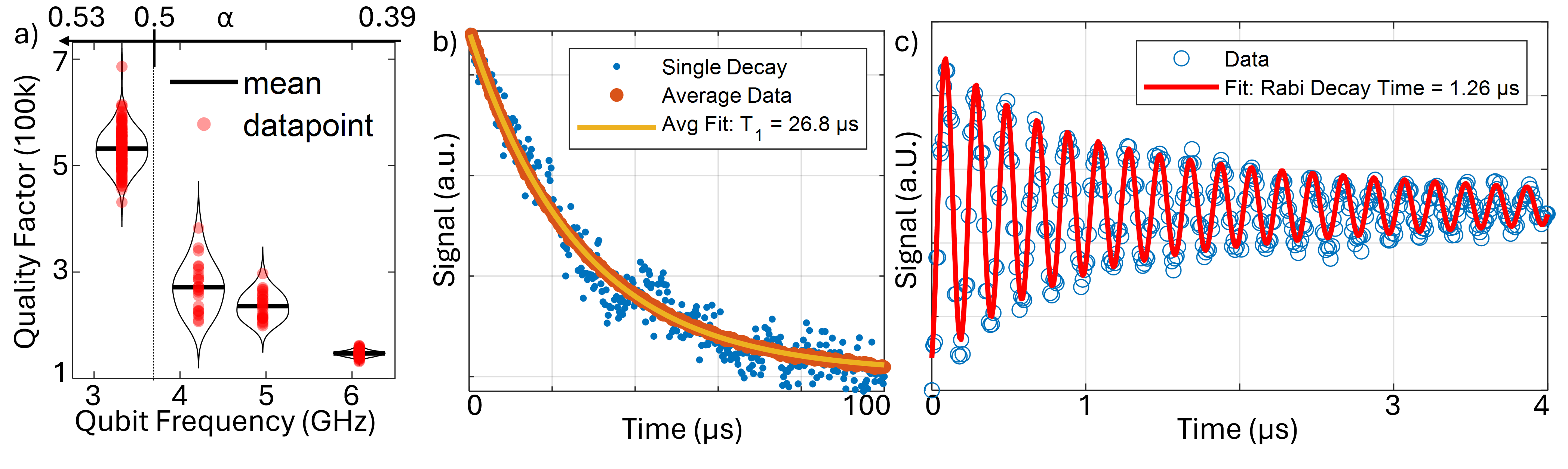}
\caption{a) Violin- and scatter-plots of quality factor measurements at the potential symmetry point $\Phi_T=0$
on sample A at different qubit frequencies,
corresponding to different $\alpha$-values at the cross-over into the double well regime ($\alpha >0.5$). The mean quality-factors (from left to right) correspond to $T_1$ of 26.8 \textmu s, 10.2 \textmu s, 7.57 \textmu s and 3.9 \textmu s. b) $T_1$-decay traces corresponding to the left-most violin. A single example trace is shown in blue and the average of all 150 traces taken over 8 hours is shown in red, with an exponential fit to the average trace in yellow.  c) Rabi oscillations measured at \fq=3.22 GHz.}
\label{fig:Qfactor&TLS}
\end{figure*}
 Figures~\ref{fig:sample}b) and c) show a photograph of our gap-tunable flux qubit sample and its circuit schematic. It features two junctions of nominally equal critical current that are connected in a gradiometric loop via two smaller-area junctions whose critical currents are reduced by a factor $\alpha_\mathrm{max}/2$.\\
The $\alpha$-junctions act as a DC-SQUID whose critical current can be controlled by an applied homogenous magnetic flux $\Phi_B$, effectively controlling the $\alpha$ factor in the qubit potential which is given by
\begin{equation}\label{eq:pot}
    U(\varphi) = -2 E_J \cos(\varphi) - \alpha\left(\Phi_B   \right)E_J \cos\left( 2\pi \frac{\Phi_T}{\Phi_0} - 2\varphi \right),
\end{equation}
with $E_J$ being the Josephson energy of one large junction and $\alpha(\Phi_B)=\alpha_{max}\cos(\pi\Phi_B/\Phi_0)$. 
Design values are chosen so that both the single-well ($\alpha<0.5$) and the double-well ($\alpha>0.5$) regime can be reached, which is illustrated in Fig.\ref{fig:results} d).  The asymmetry of the qubit potential depends on the difference flux $\Phi_T$ in the two loop branches which is controlled by a current-dividing flux bias line shown in Figs.~\ref{fig:sample}b) and c).\\
Figure~\ref{fig:results}c) shows a numerical simulation of the qubit frequency as a function of the applied flux biases, where stars and triangles respectively indicate the lowest ($\Phi_B=1\Phi_0$) and highest ($\Phi_B=2\Phi_0$) qubit frequency for a symmetric qubit potential ($\Phi_T=0$). The broad control over the qubit potential allows access to \fq~ tunability ranging from 3 MHz to 21 GHz. \\
The qubit population is controlled with resonant microwave pulses sent via the transmission line, while its state is read out by measuring the dispersive resonance shift of a capacitively coupled $\lambda/2$ readout-resonator.\\
The qubit samples are fabricated from aluminium on a sapphire substrate, using optical lithography and dry etching to pattern resonators and the qubit capacitor. The bias lines are deposited together with the Al-AlOx-Al junctions in a successive e-beam lithography step to ensure precise alignment. This reduces unwanted bias-line crosstalk. The junctions are made in a three-angle shadow evaporation process using a Dolan bridge\cite{Bilmes_bandage} which avoids the formation of unwanted stray junctions, see App. \ref{app:Fabrication} for further details. Here, we present data from two samples, whose design parameters are detailed in App. \ref{app:sampleDetails}. All measurements were performed at Millikelvin temperature in a dilution refrigerator, whose setup is detailed in App. \ref{app:setup}.\\

\section{results} \label{sec:measurement}
To characterize the qubit tunability and the cross-talk between the two flux lines, we observe the dispersive shift of the readout resonator as a function of both applied flux biases. For this, a fast method is to only measure the transmission $S_{21}$ at a fixed probe frequency $f_\mathrm{probe}$\cite{quintana2017phD}. As illustrated in Fig.\ref{fig:results}a), the transmitted signal is minimal when the readout resonator resonance $f_{\mathrm{res}}$ equals the probe frequency.
Since $f_{\mathrm{res}}$ depends on the qubit resonance frequency, this minimum indicates that the qubit was tuned to a certain frequency which depends on the chosen $f_\mathrm{probe}$.
The inside of the tear-shaped region in Fig.\ref{fig:results}b) thus corresponds to flux bias combinations where the qubit has lowest resonance frequencies.
Due to crosstalk of the flux-bias lines, the uncalibrated measurement (top panel in Fig.\ref{fig:results}b) is skewed. The lower panel shows the same measurement when the crosstalk is compensated by balancing the flux bias currents. Along the green line in the panels, the qubit potential has zero tilt $\Phi_T=0$. Points of lowest and highest qubit frequency along this line are indicated by a star and triangle as in Fig.~\ref{fig:results}c).\\
After calibration, we measure the qubit energy relaxation time $T_1$ for a symmetric potential at different qubit frequencies by tuning the size of the potential barrier via the flux \phib.
Figure~\ref{fig:Qfactor&TLS} a) shows that the qubit reaches an average quality factor $Q=2\pi \fq\,T_1$ up to 530k at an operation frequency of 3.32 GHz, which corresponds to $T_1\approx25$\,\textmu s. For this point, $T_1$ decay traces and Rabi oscillations are shown in Fig.~\ref{fig:results}b) and c). We observe an increase in qubit coherence at lower qubit frequencies where the qubit potential crosses over into the double-well regime for $\alpha>0.5$. Most of this trend is explained by a combination of readout-resonator induced Purcell-loss and limitaton by ohmic charge noise, which has been shown to dominate the energy-relaxation of C-shunted flux qubits in this frequency range\cite{Yan2016}(see App.\ref{app:energyrelaxation} for details). The quality factor is comparable to that of transmon qubits (Q $\sim$ 600k)\cite{Bilmes_bandage} and quarton qubits (Q $\sim$ 500k )\cite{kreuzer} fabricated in the same facility. To pinpoint the exact origin of the qubit decoherence, further experiments on more samples will be required to average over fabrication variations and to separate the contributions of different decoherence mechanisms in this type of flux qubit.\\

\begin{figure}[t]
\centering
\includegraphics[width=\columnwidth]{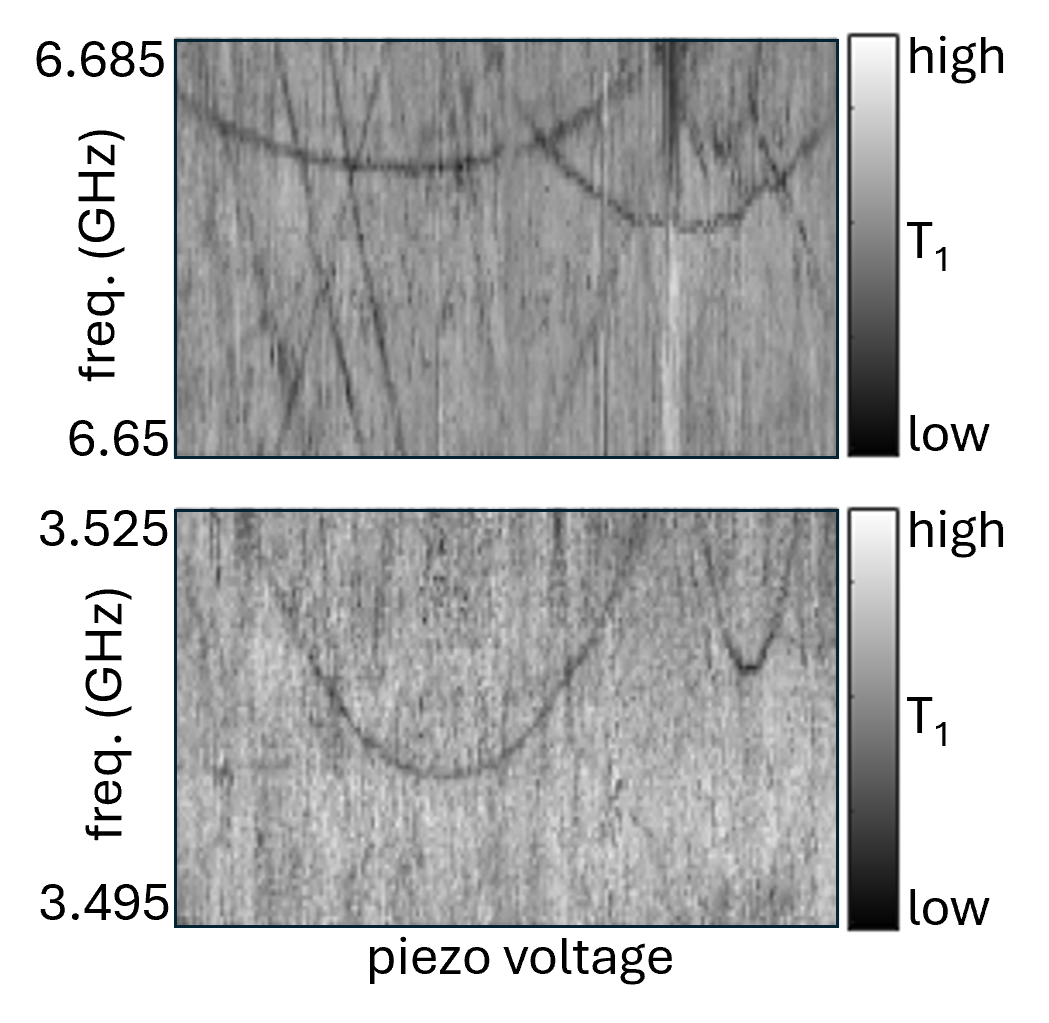}
\caption{TLS-spectra in dependence of applied mechanical strain applied via a piezo-electric element. Dark lines indicate a drop in $T_1$ caused by resonant TLS defects. Spectra were taken on sample B with similar strain ranges for different frequency intervals, using a 5~$\mu$s swap pulse. }
\label{fig:TLS_spectroscopy}
\end{figure}

The qubit's extended tunability makes it especially attractive to investigate the TLS spectral density in a wide frequency range. A powerful method to study origins of decoherence is qubit swap spectroscopy~\cite{barends13,Lisenfeld2019}, which reveals parasitic circuit modes and resonances of individual strongly-coupled TLS defects by minima in the frequency-dependent qubit $T_1$ time. With the ability to tune TLS by applied mechanical strain or DC-electric fields, one can measure spectral TLS densities and characterize the TLS' individual properties such as their coherence and dipole moment\cite{TLSosborn,Lisenfeld2016}.\\
Figure~\ref{fig:TLS_spectroscopy} shows examples of TLS spectra that were acquired with the qubit operating in the single- or double-well regime. Both spectra span a similar strain range, while a larger number of TLS are visible at higher qubit frequencies. To extract the frequency-dependent TLS density of states, it will be necessary to take into account the variation in TLS detectability due to the frequency-dependence of the qubit's $T_1$ time and TLS-qubit coupling strength that scales with $V_\mathrm{rms}\propto \sqrt{h\fq / 2C}$.\\
This first strain-tuned TLS spectroscopy in a flux qubit demonstrates TLS-detection in a frequency range spanning almost one octave. It furthermore showcases the unique combination of fast, wide, and coherent frequency tuning our design provides. We note that the experimentally demonstrated tuning range is still a fraction of what is theoretically possible. In future experiments, using an optimized experimental setup, qubit operation spanning close to four decades (3 MHz - 21 GHz) could be achieved.


\section{\label{sec:intro}Conclusion and outlook}

In this work, we have presented a gap-tunable, capacitively shunted flux qubit featuring local flux biasing and a half-gradiometric geometry. We report coherence times up to $T_1$=25\textmu s (Q=500k) and demonstrate qubit frequency tunability spanning nearly an octave, with a close to four decade range theoretically possible. This design combines the advantages of high coherence with fast and wide frequency control, making it a promising platform for quantum material research and quantum information experiments.\\
As a model application, we have demonstrated strain-tuned TLS spectroscopy, showcasing the capability to probe TLS defects across a wide frequency range and in both the single- and double-well qubit regimes. These features are particularly valuable for future investigations into TLS density of states and defect classification schemes.\\
Beyond TLS spectroscopy, the combination of good coherence and broad tunability renders this design attractive for a wide range of applications. With advances in hybrid architectures\cite{hybrid_Zhu2011,hybrid_reagor_2016,hybrid_mirul_2023}, gap-tunable flux qubits could see use as intermediaries between high-frequency qubits and low-frequency quantum memories, where their wide tuning range could mitigate frequency crowding. The capability to quickly transition between single-well and double-well potential shapes enables alternative qubit operation schemes\cite{poletto2009coherent} and provides a unique testbed for studying decoherence mechanisms in distinct potential landscapes. The ability to couple via flux\cite{coupling_vanDerPloeg2007} and to implement strong ZZ-type interactions through barrier biasing\cite{Zhu2010} further underlines their versatility as building blocks for multi-qubit systems.\\

\section*{Acknowledgments}
We thank Alexander Bilmes and Hannes Rotzinger for fruitful discussions, as well as for their contributions to the experimental setup and qubit fabrication. We thank Lucas Radtke and Silvia Diewald for their contributions to the fabrication. We are grateful for the clean room facilities provided for the fabrication by the Nanostructure Service Laboratory (NSL) at KIT. We acknowledge funding from the Baden-Württemberg Stiftung gGmbH and Google. The funders played no role in study design, data collection, analysis and interpretation of data, or the writing of this manuscript.

\bibliography{cshuntbib_truncated}

\appendix

\section{Experimental Setup}
\label{app:setup}
\begin{figure}[t]
\centering
\includegraphics[width=\columnwidth]{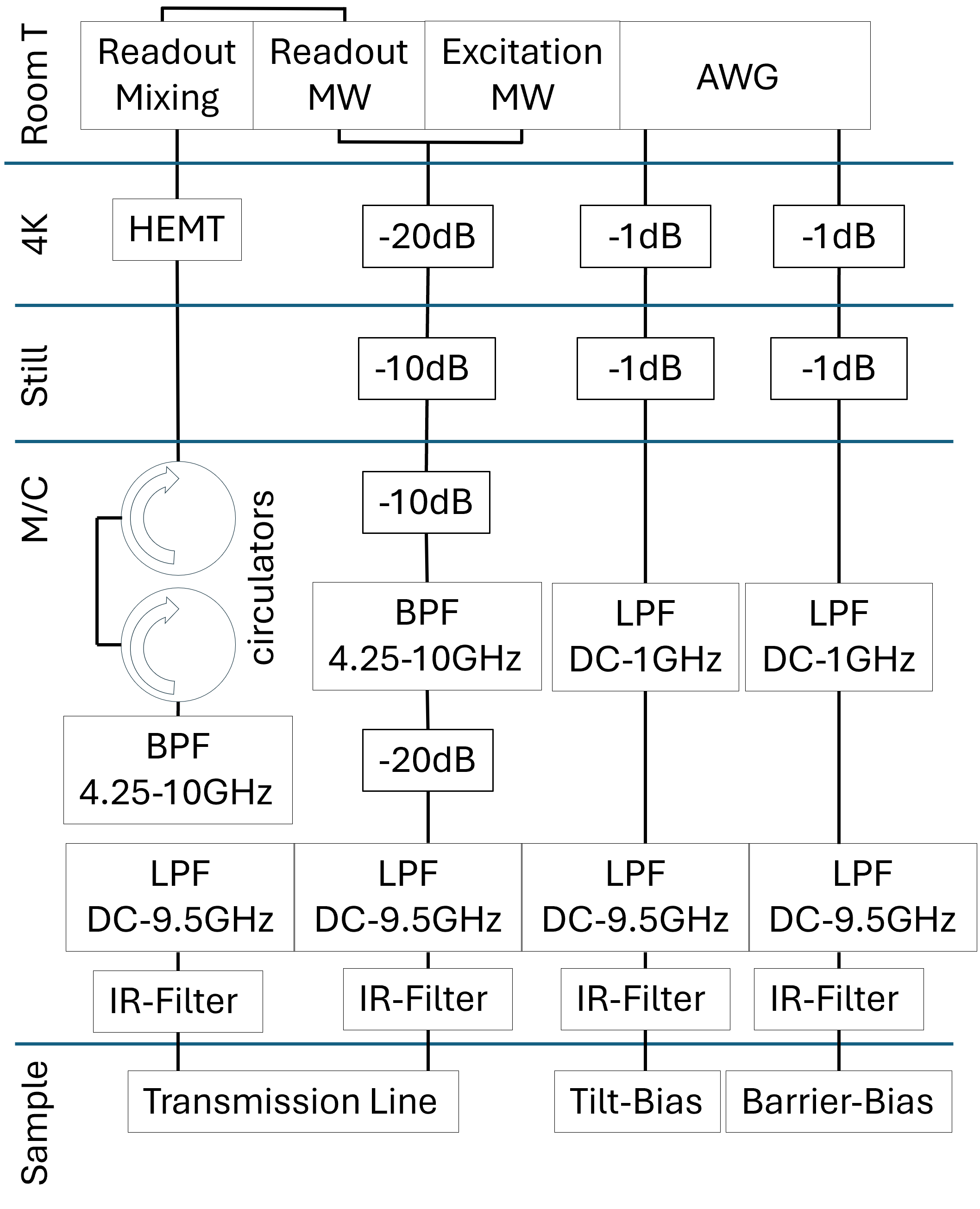}
\caption{Setup schematic of the coaxial wiring, attenuation and filtering inside the cryostat. Legend: BPF=band-pass filter, LPF=low-pass filter, IR-Filter=infrared filter, MW=microwave source, AWG=arbitrary waveguide generator, HEMT=high mobility electron transistor. }
\label{fig:setup}
\end{figure}

Figure \ref{fig:setup} shows a schematic of the attenuation and filtering setup used to minimize thermal noise and environmental radiation reaching the sample, which is located at the base stage of a dilution refrigerator (T$\approx$25mK).\\
On the input line, we apply a total of 60dB of attenuation, divided across the 4K, still, and mixing chamber (M/C) stages. This staged attenuation thermalizes incoming signals and suppresses room-temperature Johnson-Nyquist noise. Additional low-pass and band-pass filters (LPF and BPF, respectively) are placed at the base stage to reject out-of-band noise and suppress higher harmonics. On the output side, the signal passes through a pair of circulators and a HEMT amplifier at the 4K stage, followed by further amplification at room temperature. The circulators prevent amplifier noise from reaching the qubit.\\
The local qubit flux-lines are controlled via pulses generated by an arbitrary waveguide generator (AWG). On these lines, lower attenuation and 1GHz bandwidth filtering enable the fast and wide qubit flux-pulses necessary for qubit swap-spectroscopy, while preventing noise at the qubit frequency from reaching the sample.\\
In choosing the bias-line attenuation, a trade-off between tuning-range and noise suppression must be made, as highly attenuated bias lines become more prone to heating with applied bias-currents. Here, we applied minimal attenuation to explore a wide bias-flux range. Yet in an optimized setup, one should add attenuation to suppress further noise, while still accessing a $\Phi_T,\Phi_B$-range of 2$\Phi_0$ and 1 $\Phi_0$, respectively. 

\section{Sample Details} \label{app:sampleDetails}
\begin{figure}[t]
\centering
\includegraphics[width=\columnwidth]{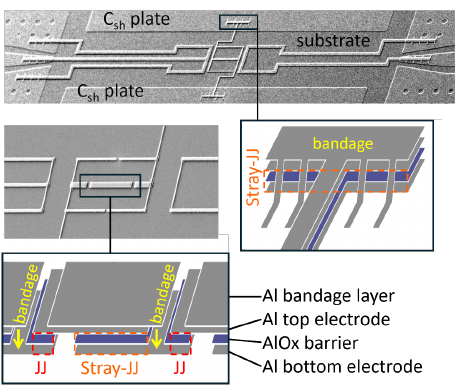}
\caption{a),b) SEM-images of samples A and B respectively, showing the bias-line and circuit layout. Inlays illustrate how an in-situ deposited bandage layer is used to avoid stray-junctions and contact the junction-circuit to the coplanar shunt-capacitor.
}
\label{fig:fabdetails}
\end{figure}

Table \ref{tab:samples} shows device parameters for the studied samples. The qubit parameters are defined with respect to the one-dimensional effective Hamiltonian\cite{Yan2016}
\begin{equation}
H = \frac{1}{2}E_C \hat{n}^2 - E_J \left[ 2\cos(\hat{\varphi}) + \alpha \cos(2\pi \frac{\Phi_T}{\Phi_0} + 2\hat{\varphi}) \right],
\end{equation}
where $\hat{n}$ is the Cooper-pair number operator and $E_C=e^2/(C_{sh}+\alpha_{max}C+C/2)$ is the effective charging energy, with C being the junction capacitance of one large junction, $C_{sh}$ the shunt capacitance and $\alpha_{max}$ the junction critical current ratio. The critical current of one small junction is by a factor $\alpha_{\mathrm{max}}/2$ smaller than that of one large junction, which results in the tunable $\alpha$-factor  $\alpha(\Phi_B)=\alpha_{max}\cos(\pi\Phi_B/\Phi_0)$. $E_J$ is the Josephson energy of one large junction, which we measure indirectly by the room temperature resistance of identically fabricated test-junctions. $\alpha_{\mathrm{max}}$ is determined in the same way, and $E_C$ is estimated via the SEM-measured junction area. The readout resonator resonance frequency \fr~ and the qubit-resonator coupling strength $g$ are measured during experiment.\\
\begin{table}[h]
\centering
\caption{Device parameters for samples A and B.}
\begin{tabular}{lcccccc}
\hline
 & $\alpha_{\text{max}}$ & $E_J$ (GHz) & $E_C$ (GHz) & $C_{\text{sh}}$ (fF) & $g$ (MHz)&$f_{\text{res}}$  \\
\hline
A & 0.85 & 164 & 0.5 & 51  & 75 & 7.662  \\
B & 0.55 & 61 & 0.57 & 51 & 62 & 7.615\\
\hline 
\end{tabular}
\label{tab:samples}
\end{table}

\section{Fabrication details}\label{app:Fabrication}
Qubits are fabricated using a three-angle Al deposition procedure, which defines the bias-lines together with the junction circuit, as illustrated in Fig.\ref{fig:fabdetails}. In a first step, the junctions are deposited using a two-angle Niemeyer-Dolan technique. Then, unwanted oxide layers are removed and an aluminium bandage layer is deposited, which connects the junction's top electrode directly to the plate capacitor (right inset). This process ensures that the unwanted stray junction (see bottom inset of Fig.\ref{fig:fabdetails}) is shorted and excessive loss from TLS in its tunnel barrier is avoided\cite{Lisenfeld2019}. This method was developed for transmon qubits by Bilmes \etal\cite{Bilmes_bandage}.

\section{Energy Relaxation}\label{app:energyrelaxation}
\begin{figure}[t]
\centering
\includegraphics[width=\columnwidth]{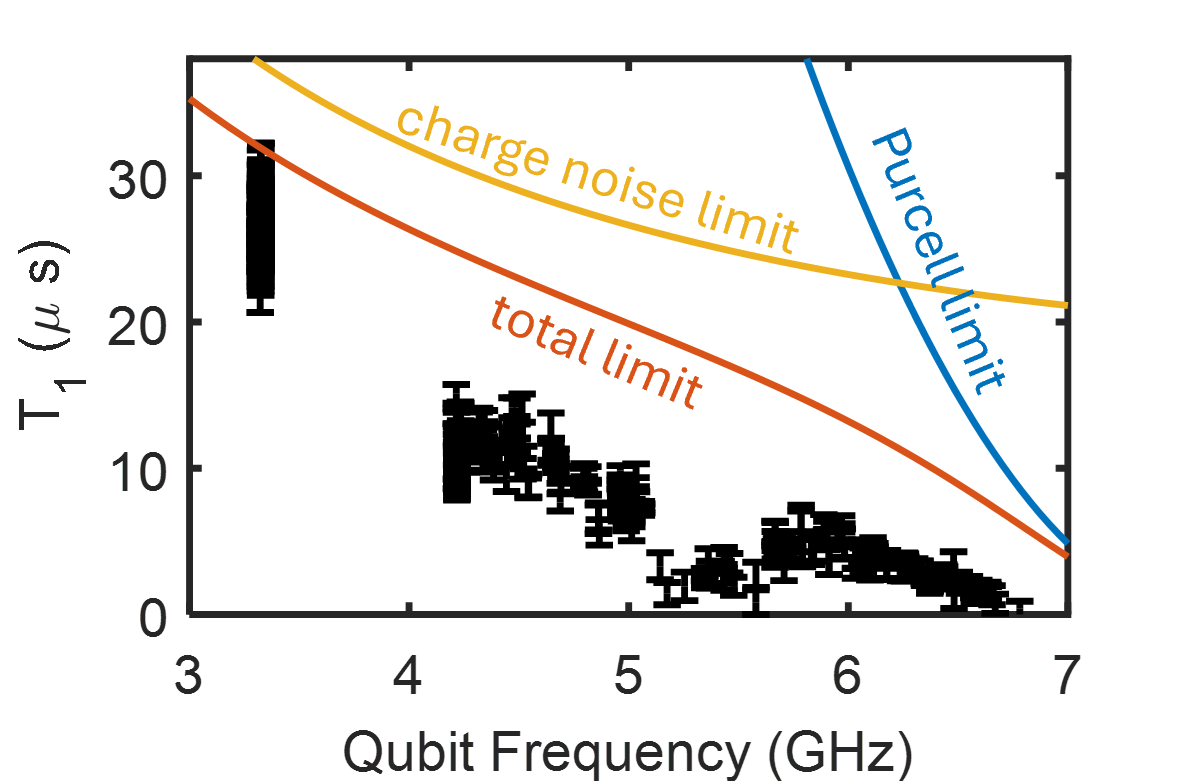}
\caption{T1 measurements (black) compared to the dominant energy relaxation limits (see text).}
\label{fig:T1LimitsApp}
\end{figure}

In the intermediate frequency range ($\sim$3-9GHz), energy relaxation of C-shunted flux qubits has been shown to be dominated by a combination of readout-resonator induced Purcell-loss and ohmic charge noise\cite{Yan2016}. The Purcell-induced limit is given by 
\begin{equation}
T_{1}^{\mathrm{Purcell}} = \frac{(2\pi)^2 (f_{\mathrm{res}} - f_{\mathrm{qubit}})^2}{g^2 \kappa},
\end{equation}
where \( f_{\mathrm{qubit}} \) is the qubit frequency, \( f_{\mathrm{res}} \) is the resonator frequency, \( g \) is the coupling strength between the qubit and resonator, and \( \kappa \) is the resonator linewidth (i.e., its energy decay rate).\\ 
The limit imposed by charge noise is calculated using Fermi's golden rule:
\begin{equation}
T_{1}^{\mathrm{charge}} = \left(\frac{1}{\hbar^2}\left| \langle 1 | \partial H / \partial Q | 0 \rangle \right|^2 S_Q(f_{\mathrm{qubit}})\right)^{-1},
\end{equation}
where \( \langle 1 | \partial H / \partial Q | 0 \rangle \) is the matrix element of the charge operator between the ground and first excited state, and \( S_Q(f_{\mathrm{qubit}}) \) is the spectral density of induced charge noise evaluated at the qubit transition frequency.  Following the calculation outlined by Yan \textit{et al.}~\cite{Yan2016}, we evaluate the charge matrix element analytically to be
\begin{equation}
\langle 1 | \partial H / \partial Q | 0 \rangle=\frac{\mathrm{n}_\mathrm{z}E_C}{e},
\end{equation}
where $\mathrm{n}_\mathrm{z}=\left(E_J/4E_C\right)^{1/4}$ is the quantum ground-state uncertainty in Cooper-pair number.\\
In the quantum regime ($hf\gg k_B T$), environmental charge noise at frequency $f$ and temperature $T$ is described by the spectral density
\begin{equation}
    S_Q(f)=C_g^2 \mathrm{Re}(Z) hf \coth\frac{hf}{2k_BT},
\end{equation}
with $Z$ being the impedance of the environment and $C_g$ the gate capacitance between environment and qubit\cite{QuantumNoiseIntroduction}.\\
Figure \ref{fig:T1LimitsApp} compares $T_1$ data measured at different qubit frequencies to the $T_1$-limits imposed by these channels, assuming Re(Z)=50\,$\Omega$ environment given by the flux bias-lines and an effective gate capacitance $C_g=0.22$\,fF. The relevant qubit parameters are found in table \ref{tab:samples}. The total $T_1$-limit $T_1^{total}=1/(1/T_{1}^{\mathrm{charge}}+1/T_{1}^{\mathrm{Purcell}})$ explains the observed overall increase in T1 towards lower qubit frequencies. Local dips in the $T_1$ time are likely caused by strongly coupled TLS.\\

\section*{Data Availability}
Data are available upon reasonable request.

\section*{Competing Interests}
All authors declare no financial or non-financial competing interests.

\section*{Funding Statement}
Funding was granted to J.L. by Google and the Baden-Württemberg Stiftung gGmbH. The funders played no role in study design, data collection, analysis and interpretation of data, or the writing of this manuscript.

\section*{Author Contributions}
B.B. designed, fabricated and measured the qubits, analyzed the data and wrote the manuscript text.
A.H. contributed to the fabrication procedures.
E.D. contributed to qubit measurement and manuscript revision.
A.U. contributed the laboratory, funding, and reviewed the manuscript.
J.L. devised the study idea, supervised the entire process and reviewed the manuscript.

\end{document}